\documentclass[aps,pre,a4paper,reprint,showpacs,notitlepage]{revtex4-1}
\usepackage{amsmath,amssymb}
\usepackage{graphicx}
\usepackage{bm}
\usepackage[usenames]{color}

\begin{document}

\title{Statistical--mechanical Analysis of Linear Programming Relaxation for Combinatorial Optimization Problems}
\date{\today}

\author{Satoshi Takabe}
\email{E-mail: s{\_}takabe@huku.c.u-tokyo.ac.jp}
\author{Koji Hukushima}
\affiliation{Graduate School of Arts and Sciences, The University of Tokyo, 3-8-1 Komaba, Meguro-ku, Tokyo 153-8902, Japan}

\begin{abstract}
Typical behavior of the linear programming (LP) problem is studied as a relaxation of the minimum vertex cover,
 a type of integer programming (IP) problem.
 A lattice-gas model on the Erd\"os-R\'enyi random graphs of $\alpha$-uniform hyperedges
 is proposed to express both the LP and IP problems of the min-VC in  the
 common statistical-mechanical model with a one-parameter family. 
Statistical-mechanical analyses reveal for $\alpha=2$ that the LP optimal solution is
 typically equal to that given by the IP
 below the critical average degree $c=e$ in the thermodynamic limit.
 The critical threshold for good accuracy of the relaxation extends the mathematical result
 $c=1$, and coincides with the replica symmetry-breaking threshold of the IP.
 The LP relaxation for the minimum hitting sets with $\alpha\geq 3$, minimum vertex covers on $\alpha$-uniform random graphs, is also studied.
 Analytic and numerical results strongly suggest that the LP relaxation fails to estimate optimal values above the
 critical average degree $c=e/(\alpha-1)$ where the replica symmetry is broken.
\end{abstract}
\pacs{75.10.Nr, 02.60.Pn, 05.20.-y, 89.70.Eg}

\maketitle

\section{Introduction}\label{sec1}
 Relaxation for discrete optimization problems is a basic and generic
 strategy to solve them 
 approximately.
 Using relaxation techniques  by which a part of an optimization problem is
 modified, we substitute easy problems for hard problems to solve.
 A striking example is a relaxation for integer programming (IP) problems.
 Although  the IP problem is generally NP-hard, the relaxed linear programming (LP) problem belongs to the class of P~\cite{elip}.
 This fact demonstrates  that the LP relaxation enables us to approximate the IP problem in polynomial time.
 The technique is applied to various practical optimizations such as vehicle routing~\cite{vr}, scheduling~\cite{acs}, and
 Boolean compressed sensing~\cite{grt}.

In this relaxation strategy, evaluating the performance 
 of approximations is an important issue both for worst-case and average-case analysis. 
 With improvement of mathematical techniques, worst-case analyses have been strongly advanced in theoretical computer science.
 The relaxation plays a key role in the construction of constant-factor-approximation algorithms for combinatorial optimization problems~\cite{vv}.
 Another attractive issue is average-case behavior of approximations for randomized optimization problems.
 It provides not only prediction of the performance of approximations but also typical hardness of optimizations.
 Analytical studies of greedy algorithms reveal average properties of problems and their intrinsic structures~\cite{fie}.
 It is still challenging, however, to study the typical behavior of relaxation analytically.
 
 Typical hardness of the optimization problems also has attracted physicists' interests
 because it is described using a type of phase transition in statistical mechanics.
 With the development of the spin-glass theory since the 1970s~\cite{sg},
 a mean-field picture with replica theory 
 has been established.
 The spin-glass techniques  were then  applied to many optimization problems.
 The picture of phase transitions breaking a replica symmetry (RS) is associated with the typical hardness of optimizations~\cite{fu,22}.
Among them, the minimum vertex cover (min-VC) has also been studied as a good
 example to which the spin-glass theory is applied. It is a well-known
 NP-hard combinatorial optimization problem defined on a graph. 
 Various types of exact or approximation algorithms
 such as a leaf removal (LR)~\cite{ks} are proposed.  The difficulty of
 approximation has been studied by computer scientists~\cite{khot}.  
  In the  statistical--mechanical view, 
  the average-case properties have been studied extensively in terms of
  phase transition~\cite{wh3}. For instance,  
  mean-field analyses of the min-VC on 
  random graphs
 conclude that 
 replica symmetry breaking (RSB) occurs at a critical average degree~\cite{wh1,zh,RR}.
 Typical  behavior of the LR 
 and its variants are also studied in solving the min-VC approximately~\cite{bg,lr0}.
 They strongly associate the typical hardness in approximation with the
 mean-field picture of the RSB transition. 
 Recently, average properties of the minimum hitting set (min-HS), the
 min-VC on hypergraphs, are also analyzed~\cite{th1,luc}.
 While the min-HS involves the multi-body interactions from the view of
  statistical mechanics, 
  it is 
  suggested that
  the goodness of the LR algorithm is characterized by the phase
 transition in the spin-glass theory.
 
  The non-trivial relation between the replica symmetry and the typical
 hardness in approximation is also suggested in the case of continuous
 relaxation by physicists~\cite{inoue,ferfon}. 
 They studied continuous relaxation with a spherical constraint, which 
 changes optimization problems to NP-hard quadratically
 constrained programming problems.
 Although it is still difficult to solve the relaxed problems, these studies indicate the existence of
  the  typical tightness of relaxation techniques.
 It is of interest whether the relation holds in the case of polynomially solvable relaxation such as the LP relaxation.
 In theoretical computer science, mathematical analysis of the LP
  relaxation for min-VCs with weights following an exponential distribution
 is performed~\cite{shah}.
 Such analyses revealed that the LP relaxation is closely related to the belief propagation in statistical physics and
 it is asymptotically tight if the belief propagation can converge with high probability.
 Recently, the typical behavior of the LP relaxation for the unweighted min-VCs is
 studied numerically~\cite{dh}, suggesting that a threshold of
 good/wrong approximation is close to the RS/RSB one  and that it is well above a
 mathematical prediction. 
 In our previous letter~\cite{th2}, we  proposed a statistical--mechanical
 analysis of the LP relaxation and showed that these two thresholds are
 coincident. 
These results  constitute a demonstration that the LP relaxation typically
 approximates an NP-hard problem with good accuracy. 

 As described in this paper,
we study the typical behavior of the LP relaxation for the min-VCs
defined on $\alpha$-uniform hypergraphs using statistical--mechanical
techniques.
The min-VC with $\alpha=2$ has 
a novel property called half integrality,  
which enables us to reduce the continuous degree of freedom in the LP to
 three states. Consequently, 
 a three-state lattice-gas model called an LP--IP model is introduced for
 studying the LP relaxation of  the min-VC.
 Statistical--mechanical analysis derives  successfully 
an analytical threshold of the typical hardness of the LP relaxation,
 which coincides with the RS/RSB transition of the original  min-VC.
 Although  a brief report on the LP--IP model based on the replica method has
 already been published~\cite{th2}, this paper presents the full
 details 
of statistical--mechanical analyses of the LP relaxation for min-VCs
 including the analysis of the cavity method.
 Additionally, we discuss  the LP relaxation for 
 the min-HS 
to examine whether its typical hardness is associated with the RS/RSB
 transition. 
Because the min-HS, unfortunately, has no half-integrality,
the LP--IP model does not completely capture the LP relaxation of the
min-HS but still provides an interesting feature on  the 
stability of integral solutions against a perturbation toward continuous values.
 
 This paper is organized as follows. 
 In the following section, we define the min-VC and its LP relaxation.
 To investigate randomized problems, random graphs and their useful properties are also introduced.
 We explain the definition of average-case properties over random graphs
 and the typical behavior of the LP relaxation.
 In section III we propose the LP--IP model  and present details of the analysis using the replica method.
 The model with three-state  Ising spins includes the min-VC. 
 It also includes LP-relaxed solutions as specific limits in a model parameter.
 By choosing 
 the parameter in the model appropriately, we obtain three 
  RS
 solutions for ground states of the model.
 We also devote some discussion to their stability.
 In section IV, we present  some numerical results of the LP relaxation. 
 In the case of the min-VCs, the statistical--mechanical analysis agrees
 well with the numerical results.
 For the min-HS, however, analytical results are no longer coincident
 with the numerical results but these results suggest that the typical
 hardness of the LP relaxation is associated with the RS/RSB
 transition. 
 The last section is devoted to a summary and discussion of the results and salient implications.
 In the Appendix, an alternative cavity analysis of the LP--IP model is
 presented. 

\section{Min-VC, LP relaxation, and their randomization}\label{sec2}
\subsection{Definitions of min-VC and hypergraphs}\label{ssec21}
 Let an $\alpha$-uniform hypergraph $G=(V,E)$ be a hypergraph of which the edges connect to $\alpha$ different vertices in $V$
 without multiplicity.
 Each vertex is labeled by $i\in V=\{1,\cdots,N\}$. Each edge in $G$ is then defined as 
 $E_a=(i_1,\cdots,i_\alpha)\in E \subset V^{\alpha}\, (i_1<\cdots<i_\alpha)$,
 where $a\in\{1,\cdots,M=|E|\}$.
 We assign a binary variable $x_i$ to the $i$-th vertex.
 The vertex $i$ is called covered if $x_i=1$, and is called uncovered otherwise.

 The min-VC problem offers each edge for the constraint that
 it should connect to at least one covered vertex.
 The covered vertex set $V'$ is defined as a subset of $V$ that satisfies all constraints for edges.
 The (unweighted) min-VC problem searches for  the minimum cardinality $|V'|$ of the covered vertex set.
 As described in this paper, the minimum cover ratio $x_c(G)=|V'|/N$ on $G$ is
 studied especially in the large-$N$ limit.
 Then, it is expressed as a form of the IP problem as 
\begin{align}
&\mathrm{Minimize}\hspace{9pt} x_c^{\mathrm{IP}}(G)=N^{-1} \bm{c}^\mathrm{T}
\bm{x},\nonumber\\
&\mathrm{Subject\ to}\,\, A \bm{x}\ge \bm{1}
,\,\bm{x}\ge \bm{0}
,\,\bm{x}\in\mathbb{Z}^N, \label{eq_21}
\end{align}
 where $\bm{x}=(x_1,\cdots,x_N)^{\mathrm T}$, $\bm{c}=(1,\cdots,1)^{\mathrm T}$, and
 an $M\times N$ incident matrix $A=(a_{ij})$ is defined as
 $a_{ai_1}=\cdots=a_{ai_\alpha}=1$ if $(i_1,\cdots,i_\alpha)= E_a$ and $a_{aj}=0$ otherwise. 
The inequality holds on each element of vectors.
 Here, the min-VC problems on hypergraphs ($\alpha\ge 3$) is especially called the min-HS.
The min-VC and min-HS,  as well as other IP problems,  are difficult to solve
 exactly in their worst case.

\subsection{LP relaxation}\label{ssec22}
 The LP relaxation is a fundamental approximation for the IP problem.
 To use the LP relaxation, it is sufficient to
 replace the integral conditions $\bm{x}\in\mathbb{Z}^N$ in the IP with 
 continuous ones $\bm{x}\in\mathbb{R}^N$.
 In the case of the min-VC, the LP-relaxed problem reads
\begin{align}
&\mathrm{Minimize}\hspace{9pt} x_c^{\mathrm{LP}}(G)=N^{-1} \bm{c}^\mathrm{T}\bm{x},\nonumber\\
&\mathrm{Subject\ to}\,\, A \bm{x}\ge \bm{1}
,\,\bm{x}\ge \bm{0}
,\,\bm{x}\in\mathbb{R}^N. \label{eq_22}
\end{align}
 Although this change on degrees of freedom engenders  good feasibility of the problems,
 it might provide  optimal solutions different from the IP problems.

 From the view of computational complexity and approximation,
 it is important whether the optimums can be obtained exactly, or not, using the LP relaxation.
 The Hoffman--Kruskal theorem is a mathematical result for the LP
 relaxation~\cite{hoff}. 
 Let us consider an LP problem given as $\mathrm{Min}\,
 \bm{\tilde{c}}^\mathrm{T}\bm{x},\, \mathrm{s.t.}\, B \bm{x}\ge
 \bm{p}$ in general.
 We define a matrix $B$ as a totally unimodular matrix if all
 sub-determinants of $B$ take only $-1$, $0$, or $1$. 
 The theorem claims that the optimal value of the LP-relaxed problem is
 equal to that of the original IP problem if the matrix $B$ is a totally
 unimodular matrix and $\bm{p}$ is an integral vector. 
Because an incident matrix $A$ of a hypertree, i.e., 
 a hypergraph with no cycles, is totally unimodular,
 the theorem ensures 
 that the optimal value of the min-VC on a hypertree can be found {exactly} by the LP relaxation.
 
\subsection{Randomized min-VC}\label{ssec23}
 As  described in Sec.~\ref{sec1}, it is our goal to find a phase
 transition of the typical behavior of the LP relaxation for the randomized min-VC.
 Here, we introduce the Erd\"os-R\'enyi random graphs as a graph
 ensemble. 
 The Erd\"os--R\'enyi random graphs are generated by choosing edges from
 all pairs of $N$ vertices with probability $p$. 
 The number of edges is then expected to be $pN(N-1)/2$. The average
 degree defined by the average number of edges connected to each vertex is $p(N-1)$.
 In this paper, 
 we 
 set $p=c/N$ where $c$ is a constant average degree of $O(1)$, leading
 to a sparse random graph. 
 In the case of $\alpha$-uniform hypergraphs, the definition of the ensemble is similar to the $\alpha=2$ case.
 Each edge is set randomly with probability $c(\alpha-1)!/N^{\alpha-1}$ from every $\alpha$-tuples of vertices.
 The degree distribution then converges to the Poisson distribution with
 mean $c$ in the large $N$ limit.  
 One of the novel properties of the ensemble is to exhibit a
 bond-percolation transition at $c_p=1/(\alpha-1)$.
 If $c<c_p$, most of vertices belong to trees and a finite
 number of short cycles exist. 
 Otherwise, a giant connected component emerges.  There exists a huge 
 number of long cycles in the component. 
 Another property is called locally tree-like structure~\cite{26}. 
 The likelihood of short cycles decays as the size of graphs grows  if the average degree $c$ is constant.
 The absence of short cycles indicates that a state on a vertex is
 predictable using information related to its neighbors. 
 This structure is especially important when the cavity method is applied to a system.
 
 The min-VC problems on the Erd\"os--R\'enyi random graphs have been studied using the replica method~\cite{wh1}
 and cavity method~\cite{zh,wz} developed in the spin-glass theory. 
These studies provide an estimation of the average minimum-cover
 ratio, i.e., an optimal value averaged over random graphs in the
 thermodynamic limit, defined as
 \begin{equation}
 x_c^{\mathrm{IP}}(c)=\lim_{N\rightarrow\infty}\overline{x_c^{\mathrm{IP}}(G)}, \label{eq_23}
 \end{equation}
 where $\overline{(\cdots)}$ is an average over the Erd\"os--R\'enyi
 random graphs with $N$ vertices and the average degree $c$.
These statistical--mechanical analyses under the RS ansatz
 estimate $x_c(c)$ of the problem, including the case of
 hypergraphs, for $c<c^*=e/(\alpha-1)$ ($e=2.71\cdots$)~\cite{th1}.
 Above the threshold $c^*$, the replica symmetry is broken, which results in an
 incorrect estimation of the minimum-cover ratio.
 Aside from these studies, it was also confirmed that a polynomial-time approximation algorithm called leaf removal
 works well in the RS region~\cite{bg}.
 However, in the RSB region, this graph-removal algorithm cannot estimate $x_c$ correctly.
 A giant connected component called LR core is left.
 These results suggest that the replica symmetry in the spin-glass theory has a close relation
 to the typical behavior of an approximation algorithm~\cite{th1,luc}.

Here, we specifically examine the LP relaxation for min-VCs and min-HSs.
 The LP-relaxed average minimum-cover ratio $x^{\mathrm{LP}}_c(c)$ is
 also a valid quantity used
 to evaluate the typical behavior of the LP relaxation.
 Given that the average degree $c<c_p$, a large part of graphs consists of (hyper)trees.
 The connected component with short cycles consists of $O(\log N)$
 vertices. Therefore, it does not affect the average ratio.
 From the Hoffman--Kruskal theorem, the LP-relaxed optimal value on (hyper)trees 
 is equal to that of the original min-VC problems.
We therefore confirm that $x^{\mathrm{LP}}_c(c)=x^{\mathrm{IP}}_c(c)$ if $c<c_p$.
 Once the bond percolation occurs above $c_p$, the Hoffman--Kruskal theorem cannot be applied directly
 because a giant component with long cycles exists.
 The recent numerical study suggests that the relation $x_c^{\mathrm{LP}}(c)=x_c^{\mathrm{IP}}(c)$ 
 is correct up to $c= 2.62(17)$~\cite{dh} above the bond-percolation
 threshold $c_p=1$ in the case of min-VCs with $\alpha=2$.
 In the next section, we analytically obtain the threshold by analyzing the LP--IP model.

\section{LP--IP model}\label{sec3}
 In this section, typical behavior of the LP relaxation for min-VC
 problems is studied using the replica method.
Although it is difficult in general to analyze a model with continuous spin variables on sparse random graphs,
 a novel property called  half-integrality enables us to estimate the LP-relaxed min-VC with $\alpha=2$
 using a statistical--mechanical method.

\subsection{Half-integrality}
 By applying an appropriate transformation, the LP problem are able to map
 onto an optimization problem 
 constrained on a convex polytope or simplex.
 Then, an extreme-point solution is defined with a feasible solution located on an extreme point of the polytope.
It is sufficient to search an extreme-point solution for solving the LP
 problem  when a cost function of the problem is linear. 
 The simplex method, 
 the first useful algorithm for the LP problems, is based on this strategy~\cite{sx}.
 Although it takes exponential time in the worst case, it solves most of the problems in polynomial time.
 
 In the case in which $\alpha=2$, the LP-relaxed min-VC problems have 
 half-integrality, that is,
 all elements  of an arbitrary extreme-point solution consist of half integers~\cite{nt}.
 From this property, we define the minimum half-integral ratio,
 \begin{equation}
 p_h(G)=\frac{1}{N}\min_{\bm{x}\mbox{:}\,\mbox{optimal}}\left|\left\{i\in V\,\bigg|\,x_i=\frac{1}{2}\right\}\right|, \label{eq_24}
 \end{equation}
 on a graph $G$. 
 It results in $x_c^{\mathrm{LP}}(G)=x_c^{\mathrm{IP}}(G)$ if $p_h(G)=0$.
 Considering random graphs, the average ratio of half integers is defined as
 \begin{equation}
 p_h(c)=\lim_{N\rightarrow\infty}\overline{p_h(G)}. \label{eq_25}
 \end{equation} 
 Along with $x_c^{\mathrm{LP}}(c)$, $p_h(c)$ provides a good evaluation of
the typical behavior of the LP relaxation.
The half-integrality also enables us to analyze the LP relaxation by the
three-state Ising model with hard-core constrants as shown later .
 As described in this paper, we specifically study 
 the model by the replica method or cavity method.
 However, the LP relaxation for the min-HS ($\alpha\ge 3$)
 has no half-integrality.
 %
 In this case, we discuss the results of the model as an approximation of the LP relaxation and
 examine its validity mainly using numerical simulations.
 
\subsection{LP--IP model}\label{ssec32}
The min-VC and min-HS are represented by a hard-core lattice gas model. 
 We first transform an occupancy variable $x_i$ to a three-state Ising variable $\sigma_i\in\{-1,0,1\}$ by $\sigma_i=2x_i-1$.
 If $\sigma_i=1$, vertex $i$ is covered and $\sigma_i=0$ represents $x_i=1/2$.
The partition function of the three-state Ising model is the following.
 \begin{equation}
 \Xi(G)=\sum_{\bm{\sigma}}\exp\left(-\mu\sum_{i}\sigma_i\right)\prod_{\{i_1,\cdots,i_\alpha\}\in E}
 \theta\left(\sum_{j=i_1}^{i_\alpha}\sigma_j+\alpha-2\right). \label{eq_31}
 \end{equation} 
 Therein, $\theta(x)$ is a unit step function that takes $1$ if $x\ge 0$ and $0$ otherwise.
 Although the ground-state energy corresponds to the LP relaxed value,
the ground states of the model might differ  from the optimal extreme-point solutions.
 On graph $G_1=(\{1,2\},\{(1,2)\})$, for example, optimal extreme-point
 solutions are $(x_1,x_2)=(1,0)$ and $(0,1)$,
 but the ground states of the model~(\ref{eq_31}) include another
 solution 
 $(x_1,x_2)=(1/2,1/2)$ in addition to the correct ones, which produces  a wrong estimation of $p_h(G_1)$.
 Omitting this trivial ground state, a penalty term is introduced as
 follows; 
 \begin{align}
 \Xi_r(G)&=\sum_{\bm{\sigma}}\exp\left(-\mu\sum_{i}\sigma_i-\mu^r\sum_{i}(1-\sigma_i^2)\right)\nonumber\\
 \times&\prod_{\{i_1,\cdots,i_\alpha\}\in E}
 \theta\left(\sum_{j=i_1}^{i_\alpha}\sigma_j+\alpha-2\right). \label{eq_31a}
 \end{align} 
 The penalty term adds some cost with a constant $r\in\mathbb{R}$ to half-integral variables.
When $r$ is larger than 1, it is regarded as Ising spin constraints in the large $\mu$ limit.
Consequently, the ground states correspond to IP optimal solutions. 
This limit is defined as an IP-limit.
In the case in which  $0<r<1$ and $\mu\rightarrow \infty$, the number of
half-integers is minimized by the penalty term though the ground-state energy is equivalent
to LP-relaxed optimal values. We thus call this limit an LP-limit.
For negative $r$, the penalty terms have no influence on the
system. This three-state limit provides the same ground states obtained by 
Eq.~(\ref{eq_31}), including trivial ground states. 
We designate this effective model the LP--IP model, which enables us 
 to estimate the LP relaxation and original IP problems in the case in which
 $\alpha=2$ 
 by setting the value of $r$ appropriately.

 The average minimum cover ratios, $x_c^{\mathrm{LP}}(c)$ and
 $x_c^{\mathrm{IP}}(c)$, are the densities averaged over the random
 graphs ensemble. 
 It is our task to calculate an average free-energy density $N^{-1}\overline{\ln\Xi_r(G)}$.
The replica method and cavity method are often used to estimate the
 free-energy density directly. 
 Here, we use the replica method developed in  an earlier study~\cite{10}.
 The alternative cavity method is  presented in the Appendix, 
 where the essentially same results derived in this section are
 obtained. 

 In the replica method, we use the replica trick $\overline{\ln\Xi_r(G)}=\lim_{n\rightarrow 0}(\overline{\Xi_r(G)^n}-1)/n$.
Considering that each edge is set randomly with probability $(\alpha-1)!c/N^{\alpha-1}$,
 the average over random graphs is taken as shown below.
 \begin{align}
 \overline{\Xi_r(G)^n}&=\sum_{\bm{\sigma}}\exp\left(-\mu\sum_{a=1}^n\sum_{i}\sigma_i^a
 -\mu^r\sum_{a=1}^n\sum_{i}\left\{1-\left(\sigma_i^a\right)^2\right\}\right)\nonumber\\
 \times&\overline{\prod_{a=1}^n\prod_{\{i_1,\cdots,i_\alpha\}\in E}\theta\left(\sum_{j=i_1}^{i_\alpha}\sigma_j+\alpha-2\right)}\nonumber\\
 &=\sum_{\bm{\sigma}}\exp\left[-\mu\sum_{a,i}\sigma_i^a-\mu^r\sum_{a,i}\left\{1-\left(\sigma_i^a\right)^2\right\}-\frac{cN}{\alpha}\right.\nonumber\\
 +&\left.\frac{c}{\alpha N^{\alpha-1}}\prod_{i_1<\cdots<i_\alpha}\prod_{a=1}^n\theta\left(\xi^a+\sum_{k}\xi_k^a+\alpha-2\right)
 +O(1)\right].
 \label{eq_32}
 \end{align} 
 We introduce an order parameter of the replicated system~\cite{10} as
 \begin{equation}
 c(\vec{\xi}\,)=\frac{1}{N}\sum_i\prod_{a=1}^n\delta(\xi^a,\sigma_i^a), \label{eq_33}
 \end{equation} 
 where $\delta(\cdot,\cdot)$ is Kronecker's delta.
 Rewriting Eq. (\ref{eq_32}) 
 by using a replicated vector $\vec{\xi}$ and its frequency ratio, 
the partition function is
\begin{align}
&\overline{\Xi_r(G)^n}
\simeq\int_{\Lambda} \left(\prod_{\vec{\xi}}dc(\vec\xi\,)\right) \exp\left[N\left(-\sum_{\vec\xi}c(\vec\xi\,)\ln c(\vec\xi\,)\right.\right.\nonumber\\
-&\left.\left.\mu\sum_{\vec\xi}c(\vec\xi\,)\xi
-\mu^r\left(n-\sum_{\vec\xi}c(\vec\xi\,)\tilde\xi\right)-\frac{c}{\alpha}\right.\right.\nonumber\\
+&\left.\left.\frac{c}{\alpha}\sum_{\vec\xi,\{\overrightarrow{\xi_k}\}}
c(\vec{\xi}\,)\prod_{k=1}^{\alpha-1}c(\overrightarrow{\xi_{k}})
\prod_{a=1}^n\theta\left(\xi^a+\sum_{k}\xi_k^a+\alpha-2\right)\right)\right],  \label{eq_34}
\end{align}
 where $\xi=\sum_{a=1}^n\xi^a$, $\tilde{\xi}=\sum_{a=1}^n(\xi^a)^2$,
 and 
 \begin{equation}
\Lambda=\left\{\{c(\vec{\xi}\,)\}\bigg|\sum_{\vec{\xi}}c(\vec{\xi}\,)=1,\,c(\vec{\xi}\,)\ge 0\: (\forall \vec{\xi}\in\{\pm 1,0\}^n)\right\}.  
 \end{equation}
 Introducing a Lagrange multiplier $\lambda$ for $\sum_{\vec{\xi}}c(\vec{\xi}\,)=1$,
 we obtain saddle-point equations for $\{\vec{\xi}\}$ as follows, 
 \begin{align}
c(\vec\xi\,)&=\exp\left[-1+\lambda-\mu\xi+\mu^r\tilde\xi \right.\nonumber\\
+&\left.c\sum_{\overrightarrow{\xi_1},\cdots,\overrightarrow{\xi_{\alpha-1}}}
\prod_{k=1}^{\alpha-1}c(\overrightarrow{\xi_{k}})
\prod_{a=1}^n\theta\left(\xi^a+\sum_{k=1}^{\alpha-1}\xi_k^a+\alpha-2\right)\right]. \label{eq_35}
\end{align}
To solve these equations, 
we assume the replica symmetric ansatz
 that the order parameter depends only on $\xi$ and $\tilde{\xi}$.
 Two effective fields $h_1$ and $h_2$ are then defined as
 \begin{equation}
c(\vec\xi\,)\overset{\mathrm{RS}}{=}c(\xi,\tilde\xi)\equiv \int dP(h_1,h_2)
\frac{1}{Z^n}\exp\left(\mu h_1\xi+\mu h_2\tilde\xi\right), \label{eq_36}
\end{equation}
 where $Z=1+2 \exp(\mu h_2) \cosh(\mu h_1)$~\cite{3st}.
Then, Eq.~(\ref{eq_35}) is represented by a joint probability distribution $P(h_1,h_2)$.
 Using the fact that the numbers of $\xi^a=-1$ and $\xi^a=0$ in
 $\vec\xi$ are given respectively by $(\tilde{\xi}-\xi)/2$ and
 $n-\tilde{\xi}$,  we find the following.
   \begin{widetext}
 \begin{align}
&
\int dP(h_1,h_2)\frac{1}{Z^n}\exp(\mu h_1\xi+\mu h_2\tilde\xi)\nonumber\\
&
=\exp\left(-1+\lambda-\mu\xi+\mu^r\tilde\xi 
+c\int \prod_{k=1}^{\alpha-1}dP(h_1^{(k)},h_2^{(k)})
\left[1-\frac{\exp\left[\mu\sum_k\left(-h_1^{(k)}+h_2^{(k)}\right)\right]}{Z^{\alpha-1}}\right]^{n-\tilde{\xi}}\right.\nonumber\\
&\times\left.\left[1-\frac{\exp\left[\mu\sum_k\left(-h_1^{(k)}+h_2^{(k)}\right)\right]}{Z^{\alpha-1}}
\left\{1+\sum_k\exp\left[\mu\left(h_1^{(k)}-h_2^{(k)}\right)\right]\right\}\right]^{\frac{\tilde{\xi}-\xi}{2}}\right).
\label{eq_37}
\end{align}
 \end{widetext}
A Laplace transformation enables us to write down a self-consistent equation of $P(h_1,h_2)$,
\begin{widetext}
  \begin{align}
 P(h_1,h_2)&=\sum_{d=0}^{\infty}e^{-c}\frac{c^d}{d!}\int\prod_{i=1}^{d}\prod_{k=1}^{\alpha-1}dP(h_1^{(i,k)},h_2^{(i,k)})
 \times
 \delta\left(h_1+1+\sum_{i=1}^du_2\left(\left\{\left(h_1^{(i,k)},h_2^{(i,k)}\right)\right\};\mu\right)\right)\nonumber\\
   \times&
   \delta\left(h_2-\mu^{r-1}+\sum_{i=1}^d\left[u_1\left(\left\{\left(h_1^{(i,k)},h_2^{(i,k)}\right)\right\};\mu\right)
 -u_2\left(\left\{\left(h_1^{(i,k)},h_2^{(i,k)}\right)\right\};\mu\right)\right]\right),
     \label{eq_38}
 \end{align}
 where
\begin{equation}
u_1\left(\left\{\left(h_1^{(i,k)},h_2^{(i,k)}\right)\right\};\mu\right)
=\frac{1}{\mu}\ln\left[1-\frac{\exp\left[-\sum_{k=1}^{\alpha-1}\mu\left(h_1^{(i,k)}-h_2^{(i,k)}\right)\right]}
{\prod_{k}\left\{1+\exp\left[\mu\left(h_1^{(i,k)}+h_2^{(i,k)}\right)\right]+\exp\left[-\mu\left(h_1^{(i,k)}-h_2^{(i,k)}\right)\right]
\right\}}\right], \label{eq_39}
\end{equation}
 and
\begin{equation}
u_2\left(\left\{\left(h_1^{(i,k)},h_2^{(i,k)}\right)\right\};\mu\right)
=\frac{1}{2\mu}\ln\left[1-\frac{\exp\left[-\sum_{k=1}^{\alpha-1}\mu\left(h_1^{(i,k)}-h_2^{(i,k)}\right)\right]
\left(1+\sum_k\exp\left[\mu\left(h_1^{(i,k)}-h_2^{(i,k)}\right)\right]\right)}
{\prod_{k}\left\{1+\exp\left[\mu\left(h_1^{(i,k)}+h_2^{(i,k)}\right)\right]+\exp\left[-\mu\left(h_1^{(i,k)}-h_2^{(i,k)}\right)\right]\right\}}\right]. \label{eq_40}
\end{equation}
\end{widetext}
 Our aim is to solve this equation in $\mu\rightarrow \infty$ limit.
 The parameter $r$ has a crucial role in the limit. 
 The following three cases are characterized by the value of $r$. 

\subsection{Case 1: IP-limit $(r>1)$}
 In the case in which $r>1$, the effective field $h_2$ diverges.
 Then, the self-consistent equation of $P(h_1,\infty)$ is reduced to 
\begin{align}
&P(h_1,\infty)=\sum_{d=0}^{\infty}e^{-c}\frac{c^d}{d!}\int\prod_{i=1}^{d}dP(h_1^{(i)},\infty)\nonumber\\
\times&\delta\left(h_1+1+2\sum_i\prod_{k=1}^{\alpha-1}\theta\left(-h_1^{(k)}\right)
\mathrm{max}\left(h_1^{(1)},\cdots,h_1^{(\alpha-1)}\right)\right). \label{eq_41}
\end{align}
 This equation is equivalent to that of the original min-VC on $\alpha$-uniform hypergraphs~\cite{th1}.
 $P(h_1,\infty)$ has a sharp peak around some integral values of $h_1$
 if $\mu\gg 1$. 
 We therefore assume an integer-field ansatz that the effective field $h_1$ takes integer in $\mu\rightarrow\infty$ limit.
 Eq.~(\ref{eq_41}) is solved under this ansatz. The average minimum cover ratio is expressed as shown below.
\begin{equation}
x_c^{\mathrm{IP}}(c)=1-\left[\frac{W((\alpha-1)c)}{(\alpha-1)c}\right]^{\frac{1}{\alpha-1}}
\left[1+\frac{W((\alpha-1)c)}{\alpha}\right],
\label{eq_42}
\end{equation}
 Therein, $W(x)$ denotes the Lambert's W function defined by $W(x)e^{W(x)}=x$.
 The RS ansatz gives the correct value of $x_c^{\mathrm{IP}}(c)$ below the threshold $c^\ast=e/(\alpha-1)$.

\subsection{Case 2: LP-limit $(0<r<1)$}\label{ssec34}
\begin{figure}[!tb]
\begin{center}
 \includegraphics[trim=0 0 0 0,width=1.0\linewidth]{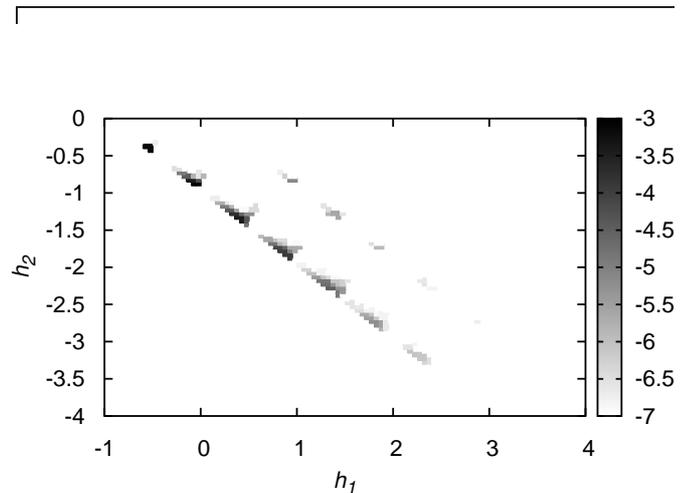}
 \caption{Saddle-point solution $P(h_1,h_2)$ of Eq.~(\ref{eq_38}) obtained using a population dynamics
 with $c=4$, $\mu=30$ and $r=0.01$.
 The fraction on each point is shown by the logarithmic gray scale.
 The number of population is $10^4$ and $10^4$ iterations are executed.
}
 \label{fig0}
\end{center}
\end{figure}

 Let us consider the case in which $0<r<1$.
 Fig.~\ref{fig0} shows a numerical solution of Eq.~(\ref{eq_38}) with $\mu=30$ obtained using the population dynamics~\cite{popdy}.
 Results show that the joint probability density $P(h_1,h_2)$ is supported on triangular parts
 located at $(h_1,h_2)=(m+l/2-1,-l/2)$  with $m,l\ge 0$ and $m,l\in\mathbb{Z}$.
 Considering that the effective fields fluctuate because of the infinitesimal penalty $\mu^{r-1}$,
 these values are represented by $(m+l/2-1+v\mu^{r-1},-l/2+w\mu^{r-1})$
 with some coefficients $v$ and $w$. 
 The numerical simulations 
 imply that the fluctuation has the following property 
\begin{equation}
w\ge 1,\quad -w+1\le v \le w-1. \label{eq_43}
\end{equation}
 This infinitesimal-field ansatz is conserved by Eq.~(\ref{eq_38}). It
is also consistent with numerical solutions obtained by the population
dynamics.  
 The joint probability distribution of the effective field is then
 decomposed into some probabilities with support on each triangle as 
\begin{align}
 P(h_1,h_2)&=\sum_{l,m=0}^{\infty}R(l,m)
\end{align}
where
\begin{align}
R(l,m)&=\int dP(h_1,h_2)\sum_{(v,w)\in \mathcal{D}} \delta\left(h_1-\left(m+\frac{l}{2}-1\right)-v\mu^{r-1}\right)\nonumber\\
\times&\delta\left(h_2+\frac{l}{2}-w\mu^{r-1}\right), \label{eq_44}
\end{align}
and 
$\mathcal{D}=\{(v,w)\in \mathbb{Z}^2|w\ge 1,\, -w+1\le v \le w-1\}$.

 A set of effective fields $(h_1,h_2)$ is distinguished using a likelihood of spin values.
 We define several regions as follows: $\mathcal{P}=\{(h_1,h_2)|\,h_2<-|h_1|\}$, 
 $\mathcal{Q}=\{(h_1,h_2)|\,h_2>h_1,\,h_1<0\}$, $\mathcal{R}=\{(h_1,h_2)|\,h_2>0,\,h_1=0\}$,
 and $\mathcal{S}=\{(h_1,h_2)|\,h_2>-h_1,\,h_1>0\}$.
 When we define 
 a set of probabilities that a spin takes $1$, $0$, and $-1$ as
 $(p_1,p_0,p_{-1})$,
 the sets in each region of $\mathcal{P}$, $\mathcal{Q}$, $\mathcal{R}$
 and $\mathcal{S}$ are $(0,1,0)$, $(0,0,1)$, $(p_\alpha,0,1-p_\alpha)$
  with $p_\alpha\in(0,1)$ and $(1,0,0)$, respectively. 
 Assuming Eq.~(\ref{eq_43}), the weights of these states read 
\begin{align}
P&=\sum_{l=2}^\infty R(l,0),\,Q=R(0,0)+R(1,0),\nonumber\\
R&=R(0,1),\,S=\sum_{m+l\ge 2,m\ge 1}R(l,m). \label{eq_45}
\end{align}
 Eq.~(\ref{eq_38}) enables us to obtain self-consistent equations as follows
 \begin{align}
P&=\sum_k e^{-c}\frac{c^k}{k!}\left\{1-(Q')^{\alpha-1}\right\}^k-Q
=e^{-c(Q')^{\alpha-1}}-Q, \nonumber\\
Q&=Q'+Q'', \nonumber\\
Q'&=\sum_k e^{-c}\frac{c^k}{k!}\left\{1-(Q')^{\alpha-1}-(\alpha-1)
(P+Q'')(Q')^{\alpha-2}\right\}^k\nonumber\\
&=\exp[-c(Q')^{\alpha-2}\{(\alpha-1)(P+Q)-(\alpha-2)Q\}]
, \nonumber\\
Q''&=\sum_ke^{-k}\frac{c^k}{k!}k(\alpha-1)(P+Q'')(Q')^{\alpha-2}\nonumber\\
&\times\left\{1-(Q')^{\alpha-1}-(\alpha-1)
(P+Q'')(Q')^{\alpha-2}\right\}^{k-1}\nonumber\\
&=c(\alpha-1)(P+Q'')(Q')^{\alpha-1}, \nonumber\\
R&=\sum_k e^{-c}\frac{c^k}{k!}k(Q')^{\alpha-1}\nonumber\\
&\times\left\{1-(Q')^{\alpha-1}-(\alpha-1)
(P+Q'')(Q')^{\alpha-2}\right\}^{k-1}\nonumber\\
&=c(Q')^\alpha, \nonumber\\
S&=1-P-Q-R, \label{eq_46}
\end{align}
 where $Q'=R(0,0)$ and $Q''=R(1,0)$.
 Substituting $X=P+Q$ and $Y=Q'$, we find
\begin{equation}
 X=\exp(-cY^{\alpha-1}),\, Y=\exp[-cY^{\alpha-2}((\alpha-1)X-(\alpha-2)Y)]. \label{eq_47}
\end{equation}
 The spin variable takes $1$ with probability $p_\alpha$ and $-1$ otherwise if $(h_1,h_2)$ is located in region $\mathcal{R}$.
 It is the third ansatz to consider the probability $p_\alpha=1/\alpha$ on $\alpha$-uniform hypergraphs.
 Then, using the solution of Eq.~(\ref{eq_47}), the LP-relaxed average minimum cover ratio reads
\begin{align}
 x_c^{\mathrm{LP}}(c)&=1-\frac{1}{2}\left[X+Y+c(\alpha-1)(X-Y)Y^{\alpha-1}\right.\nonumber\\
 &\left.+2c\frac{\alpha-1}{\alpha}Y^\alpha\right], 
 \label{eq_48}
\end{align}
and the average fraction of half integers is represented as
\begin{equation}
p_h(c)=(X-Y)(1-c(\alpha-1)Y^{\alpha-1}).
\label{eq_49}
\end{equation}
 
 For any $\alpha$, $X$ is equal to $Y$ below the average degree
 $c^\ast=e/(\alpha-1)$. 
 In this case, $X=Y=[W((\alpha-1)c)/(\alpha-1)c]^{1/(\alpha-1)}$ engenders $x_c^{\mathrm{LP}}(c)=x_c^{\mathrm{IP}}(c)$ and $p_h(c)=0$,
 which suggests that the LP relaxation typically solve the problem with high accuracy.
 However, it is apparent that $X>Y$ leads to $p_h(c)>0$.
 As presented in later sections, the LP-relaxed value is apparently below the optimal one.
 These facts reveal that a phase transition as for the typical behavior of the LP relaxation occurs
 at critical average degree $c=c^\ast$.
 In the case of $\alpha=2$, $p_h(c)$ is equivalent to the average fraction of a core generated by a leaf removal algorithm~\cite{bg}
 though it is not the case if $\alpha\ge 3$~\cite{th1}.

\begin{figure}[!tb]
\begin{center}
 \includegraphics[trim=0 0 0 0,width=1.0\linewidth]{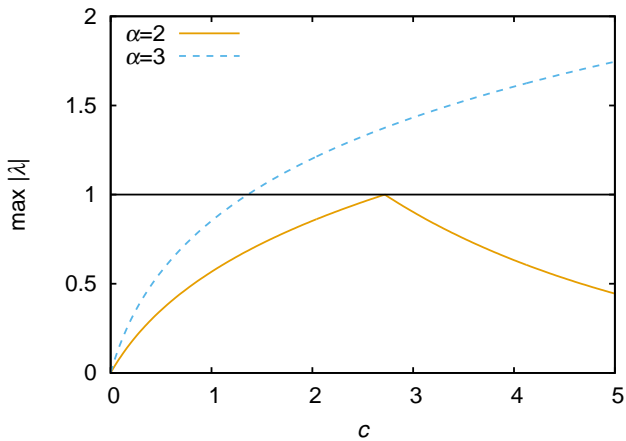}
 \caption{(Color online) Maximal absolute eigenvalue for the
 local-stability matrix as a function of the average degree $c$.
 Solid and dotted line represent the case of min-VC ($\alpha=2$) and min-HS ($\alpha=3$), respectively.
 The horizontal solid line is $\max |\Lambda(c)|=1$, above which 
 the local stability breaks.
}
 \label{figa1}
\end{center}
\end{figure}

 Here, we discuss the stability of the RS solution.
 In terms of statistical mechanics,
the convexity of the free energy called the de-Almeida and Thouless (AT) condition~\cite{at} is a reasonable
 qualifications to study its stability.
 Unfortunately, however, no method has been established to verify the
 AT condition of the models defined on finite connectivity graphs.
 As a necessary condition, we study local stability of the self-consistent equations~\cite{zma}.
A perturbation $(\delta X,\delta Y)$ added to a possible solution
  $(X,Y)$ is transformed through Eq.~(\ref{eq_47}) as
 \begin{equation}
\begin{pmatrix}
\delta X'\\
\delta Y'
\end{pmatrix}
=
\begin{pmatrix}
0& WX\\
WY& (\alpha-2)W(X-Y)
\end{pmatrix}
\begin{pmatrix}
\delta X\\
\delta Y
\end{pmatrix},\label{eq_50}
\end{equation}
where $W=-c(\alpha-1)Y^{\alpha-2}$.
  The eigenvalues of the matrix are
\begin{equation}
 \Lambda(c)=\frac{W}{2}\left[(\alpha-2)(X-Y)\pm\sqrt{(\alpha-2)^2(X-Y)^2+4XY}\right]. \label{eq_51}
\end{equation}
 The solution of Eq.~(\ref{eq_38}) is stable
 in terms of its self-consistent equations if the maximal absolute value
 of these eigenvalues is below 1. 
 In the case of the min-VC with $\alpha=2$, $\Lambda(c)$ increases below
 $c<c^\ast=e$ and reaches $1$ at $c=c^\ast$. 
 Above the threshold, however, it decreases and the RS solution remains
 stable up to $c=\infty$, as shown in Fig.~\ref{figa1}. 
 In contrast, $\Lambda(c)$ of min-HSs with $\alpha\geq 3$ increases
 monotonously. The RS solution loses its linear stability above the
 threshold. 
 This difference shows that the half-integer relaxation in our model is
 insufficient for the min-HS to describe the LP-relaxed solutions, 
 whereas the min-VC holds the half-integrality.

\subsection{Case 3: three-state limit}
For the parameter $r<0$, the penalty term does not affect on the
 system. 
 The ground states consist not only of optimal extreme-point solutions but also of other trivial ground states.
 The RS solution in this limit thus can not predict the typical behavior of the LP relaxation except for its approximate value.
 For example, the half-integral ratio $p_h$ is always positive for any
 $c$,  which is
 quite different from numerical results of the 
  LP problem shown below.

\section{Numerical simulations}
 In this section, we perform numerical simulations of two types 
 to confirm our RS analyses in the LP-limit and IP-limit.
 One is the Markov-chain Monte Carlo simulation for estimating optimal values of
 the original problems.
 We especially use the replica-exchange Monte Carlo (EMC)~\cite{90,91}
 method to accelerate equilibration of the system.
 We set 50 replicas with different values of chemical potential.
 An optimal value on each graph is evaluated by the minimum density found in at least $2^{17}$ Monte Carlo steps.
It is then averaged over $800$ random graphs with $16\mathchar`-512$
 vertices and extrapolated to $x_c^{\mathrm{IP}}$ 
 using a quadratic function of $N^{-1}$.
 The evaluated optimal values are compared to the analytical RS solutions of the LP--IP model.
 The other is LP relaxation.
 It is performed mainly to examine the validity of LP-limit solutions for both min-VCs and min-HSs.
 We generate at least 800 random graphs and solve the LP-relaxed problems by
 a revised simplex method using LP\_solve\_5.5 solver~\cite{lpsolve}.
 Especially in the case of min-VCs, the LR algorithm is executed as pretreatment
 because of accurate estimation of the half-integral ratio $p_h$.
  
\begin{figure}[!tb]
\begin{center}
 \includegraphics[trim=0 0 0 0,width=1.0\linewidth]{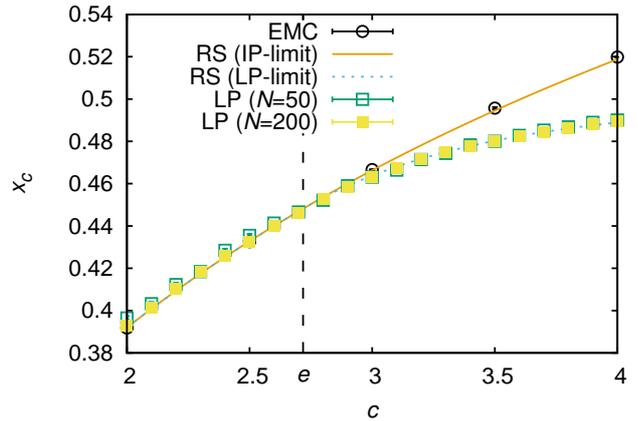}
 \caption{(Color online) The minimum-cover ratio in Erd\"os--R\'enyi random graphs with $\alpha=2$ as a function of the average degree $c$.
 Circles are numerical results given by the replica exchange Monte Carlo method.
 Square marks are numerical results obtained using the revised simplex method with vertex cardinalities $N=50$ (open)
 and $N=200$ (filled).
 These are averaged over $800$ random graphs for the Monte Carlo method and $1600$ random graphs for the LP relaxation.
 The solid and dotted lines respectively show the RS solutions in the IP-limit and the LP-limit.
 The vertical dashed line represents the critical average degree $c^\ast=e$.
}
 \label{fig1}
\end{center}
\end{figure}

 We first discuss numerical results for the optimal or approximate values of min-VCs.
 Fig.~\ref{fig1} shows optimal or approximate cover ratios obtained using the EMC and LP relaxation.
 For a relatively small average degree, it is apparent that the RS solutions and LP-relaxed numerical results
 well agree with the optimal values estimated by the EMC.
 This shows that the LP relaxation typically approximates the original problems in good accuracy in the RS phase.
 In contrast, when the average degree is above the critical threshold $c=e$, the RS solutions in the IP-limit become unstable.
 It leads to a wrong evaluation for the optimal values compared to the EMC.
 Then higher RSB solutions are necessary to estimate the optimal values exactly.
 In the case of the LP relaxation, our statistical--mechanical prediction
 still agrees with the numerical data.
 We also confirm that the LP relaxation typically fails to estimate
 the optimal values if the average degree is larger than $c^\ast$.
  The LP-relaxed approximate value of the min-VC goes to $1/2$ in the
 large $c$ limit, 
 while the optimal value of the min-VC is asymptotically close to $1$.
  
\begin{figure}[!tb]
\begin{center}
 \includegraphics[trim=0 0 0 0,width=1.0\linewidth]{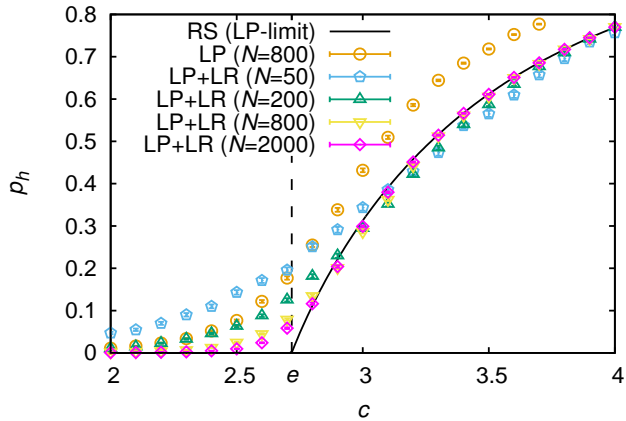}
 \caption{(Color online) Half-integral ratio $p_h$ as a function of the average degree $c$.
{Circles} denote data obtained only using the revised simplex method with 
vertex cardinality $N=800$.
 Other open marks are numerical results obtained using the simplex method after running a leaf removal algorithm
 with 
 $N=50$, $200$, $800$, and $2000$. 
 These are averaged over {$1600$} random graphs.
 The solid line represents the RS solution in the LP-limit.
 The vertical dashed line represents the critical average degree $c^\ast=e$.}
 \label{fig2}
\end{center}
\end{figure}

 Next, we specifically examine the half-integral ratio $p_h$ representing a typical property of the approximate solutions.
 In Fig.~\ref{fig2}, it is apparent that numerical data obtained using the LP relaxation is
 well above our analytic prediction.
Generally speaking, LP-relaxed problems have several optimal extreme-point solutions
 because of the existence of a leaf, a pair of vertices either of which are of degree one.
 For instance, we assume that a graph $G_2$ consists of an odd cycle and a leaf, and that
 a vertex in the cycle is connected to one in the leaf by an edge.
 Then, an LP-relaxed min-VC on $G_2$ has two solutions: one has all half-integral variables.
 The other has integral variables in the leaf.
 If one simply runs a solver, then one obtains the average ratio with half-integral variables, 
 not the minimum ratio $p_h$ predicted by the LP--IP model.
 For this reason, the discrepancy in $p_h$ arises.
 We therefore perform an LR algorithm before executing the LP relaxation, by
 which half-integral variables induced by the leafs can be avoided.  
 Fig.~\ref{fig2} shows the minimum half-integral ratio estimated using the
 procedure.
 As expected, the modified LP method reduces the number of
 half-integral variables after performing the LR algorithm.
 Therefore, this LR+LP method obtains the
 optimal extreme-point solutions and improves the numerical estimation
 of $p_h$.
 Although there remains a finite-size effect for small
 sizes and around the threshold $c\simeq e$,
 the numerical estimations are close to the analytic results with
 increasing size. 
 Our analysis correctly predicts not only the approximate value of $x_c$,
 but also the typical property of the LP relaxation.
 
\begin{figure}[!tb]
\begin{center}
 \includegraphics[trim=0 0 0 0,width=1.0\linewidth]{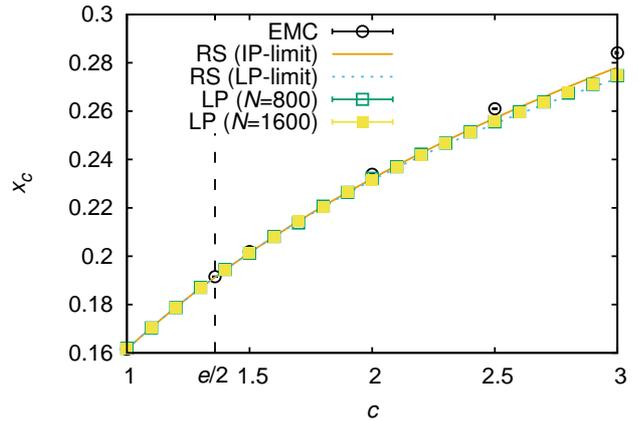}
 \caption{(Color online) Minimum-cover ratio in Erd\"os--R\'enyi random graphs with $\alpha=3$ as a function of the average degree $c$.
 Circles are numerical results given by the {replica} exchange Monte
 Carlo (EMC) method.
 Square marks denote the data obtained using the revised simplex method for vertex cardinalities of $N=800$ (open)
 and $N=1600$ (filled).
 These are averaged over $800$ random graphs.
 The solid and dotted lines respectively show the RS solutions in the IP-limit and the LP-limit.
 The vertical dashed line represents the critical average degree $c^\ast=e/2$.
}
 \label{fig3}
\end{center}
\end{figure}

 Lastly, we present the case of min-HS problems with $\alpha=3$.
 Fig~\ref{fig3} shows the optimal values estimated using the EMC and
 the approximate values obtained by the LP relaxation, together with the
 analytical results derived in the previous section. 
 All the results coincide mutually for a sufficiently small
 average degree. 
 The relaxed values, however, are markedly smaller than the optimal
 values of the original problem above the critical average degree $c=e/2$,
 where the replica symmetry of the min-HS is broken.
 We therefore confirm that the LP relaxation typically fails to approximate min-HSs in the RSB region.
 As a striking difference between min-VCs and min-HSs, we point out that the RS solutions
 in the LP-limit are also unstable above the critical threshold.
 Whereas the discrepancy 
 between the numerical LP-relaxed results and the analytic estimations
 is quite small as shown in Fig~\ref{fig3},
 it increases gradually as $c$ becomes large.
 In the large-$c$ limit, the LP-relaxed value on $\alpha$-uniform graphs converges to $1/\alpha$
 whereas the analytic solutions converge to $1/2$.
 Our result implies that the existence of the RSB region in the LP-limit
 results from the lack of half-integrality in min-HSs.
 To obtain a better analytic prediction, one must consider
 the model with more degrees of freedom, beyond the half-integrality condition.

\section{Summary and Discussion}
In this paper, we describe the details of the statistical--mechanical analysis of typical behavior of the LP relaxation.
The LP relaxation of the min-VC can be mapped onto the LP--IP model with
three-state Ising variables assisted by
 				 the novel property called
 				 half-integrality.
Three distinct ground states are derived 
by fixing a parameter $r$ of the model and taking a large field (zero temperature) limit.  
 The replica method in the spin-glass theory enables us to solve the model approximately  in these limits.
 In the IP-limit with $r>1$,
 the ground states are reduced to optimal solutions of the original
 min-VCs. 
 The ground states in the LP-limit with $0<r<1$ correspond to the
 LP-relaxed approximate solutions with minimum half-integral variables. 
 In the three-state limit with $r<0$,
 the ground states are not constructed by the extreme-point solutions
 which are unsuitable for the LP-relaxed solution. 
 The RS solution in the LP-limit is stable for the arbitrary average degree.
 Therefore, the LP-limit solution coincides with the numerical result.
 However, the RS solution in the IP-limit is unstable above
 $c^\ast=e$. In fact 
 the LP relaxation fails to approximate optimal solutions above the critical threshold.

 We also discuss the case of the min-HS, min-VCs on $\alpha$-uniform
 hypergraphs. 
 Because the min-HS has no half-integrality, the LP--IP model
 with three-state Ising is insufficient for describing the LP relaxation
 of the min-HS.
 It is, however, worth studying the LP--IP model for a half-integer
 relaxed problem toward an understanding of the LP relaxation.
 It is particularly interesting that the RS solution in the LP limit is still stable 
 below the critical threshold of the min-HS.
 This stability suggests that  the original problem is stable against addition of the
 half-integral variables to its solution. 
 Above the threshold, the analytic estimation by the RS solution in the
 LP-limit deviates 
 from the optimal value of the original problem. 
 The RS solution is simultaneously
 unstable, meaning the emergence of the RSB solutions.  
 This fact implies that the half-integer relaxed problem 
  decreases the value of the cost function from the original
 problem but it 
 is still typically difficult to solve. 
 In fact,
 LP-relaxed approximate solutions obtained using the numerical simulations
 include not only half integers but also other real values.
 These results suggest that the LP-relaxed min-HS typically fails to approximate the original problem in the RSB region.
 
One of the striking facts 
 obtained through our analysis is that, in the case of
 $\alpha=2$, 
 the minimum half-integral ratio $p_h$ 
 has the same mathematical expression as 
 the LR core~\cite{bg}. 
 It strongly suggests that
 a common graph structure is the origin of the wrong estimation in two
 different approximation algorithms,  
 which is unfortunately not identified.
 A key ingredient of the graph structure may be the core, in which there
 exist entangled odd long cycles and clustering of the optimal solutions
 occurs~\cite{Bar}. 
 In min-$K$-XORSAT, a $K$-core is also regarded as a trigger for the typical
 hardness\cite{xor2}. 
 Then, it is naively expected that some graph structures will be a cause of
 both the replica symmetry breaking and the typical hardness in 
 other combinatorial optimization problems.
 As for min-HSs, in contrast, it is an open problem whether the minimum non-integral ratio is related to the core ratio.
 In general, the relation between an emergence of some graph structures and the RS/RSB transition is thus still to be revealed.
 It is interesting to consider the RSB picture more generally from the perspective of graph topology.

 In this paper, we utilize the half-integrality for constructing the LP--IP model.
 Recently, from the view of discrete convexity, bisubmodular relaxation which is equivalent to the LP relaxation with half-integrality
 is proposed~\cite{Kor}.
 It is related closely  to an approximation tequnique previously known as the roof duality.
 It has been applied to more general approximation called generalized roof duality in optimization and inference~\cite{kahl}.
 Because variables in a relaxed problem take $\{0,1/2,1\}$, the LP--IP model and its analyses in this paper are applicable to the relaxation.
 Statistical-mechanical approaches will be of help to elucidate a typical property of these schemes theoretically.
 
 We have demonstrated statistical--mechanical analysis of the typical
 behavior of an approximation algorithm for combinatorial optimization
 problems. 
 Particularly, we emphasize  on the LP relaxation based on the
 simplex method, which searches extreme points of a polytope generated by constraints.
 We construct effective model as the LP--IP model by extending the degree of freedom of spins and adding a penalty term
 to a conventional hard-core lattice gas model for the min-VC.
 Within the framework of the LP relaxation,
 theoretical standard model is necessary for the 
relaxed problems without the half-integrality  and
also for 
other solvers such as a cutting-plane approach~\cite{sch}. 
 Another task is a statistical--mechanical study on other relaxations proposed in the literature of mathematical optimization.
 These analyses are expected to be helpful to provide conjecuters related to the average complexity of optimization problems 
 in theoretical computer science and probability theory.
 We hope that they are useful to investigate the deep relation between the spin-glass theory and optimization problems.
 
\begin{acknowledgments}
 This research was supported by Grants-in-Aid for Scientific Research
 from  MEXT, 
 Japan (Nos. 22340109, 25610102, and 25120010), for JSPS fellows (No. 15J09001),
 and by the JSPS Core-to-Core program
 ``Non-equilibrium dynamics of soft-matter and information.''
\end{acknowledgments}

\appendix
\section{Cavity analysis of the LP--IP model}\label{sec_app}
 In this appendix, we  present detailed analyses of the model discussed in
 this paper using the alternative cavity method.
 Although we explain the case of $\alpha$-uniform random hypergraphs here, it is straightforward to
 calculate more general models defined on a sparse hypergraph. 

 Using a factor graph representation $G=(V,F,E)$, the LP--IP model~(\ref{eq_31}) is represented as
\begin{align}
 \Xi_r(G)&=\sum_{\bm{\sigma}}\exp\left(-\mu\sum_{i}\sigma_i-\mu^r\sum_{i}(1-\sigma_i^2)\right)\nonumber\\
 \times&\prod_{a\in F}
 \theta\left(\sum_{j\in\partial a}\sigma_j+\alpha-2\right), \label{eq_a1}
\end{align}
 where $\partial a=\{i\in V\,|\, (a,i)\in E\}$.
 We assume that the graph is locally tree-like and that it has no degree correlations.
 By the Bethe--Peierls approximation, the likelihood that a variable on vertex $i$ takes $\sigma$ is 
\begin{equation}
P_i(\sigma)\simeq \frac{1}{Z_i}\exp(-\mu\sigma-\mu^{r}\delta_{\sigma,0})
P_{a\backslash i}(\sigma)
,\label{eq_a2}
\end{equation}
 where $P_{a\backslash i}(\sigma)$ is the marginal probability of $\bm{\sigma}_{\partial a\backslash i}$
 under the condition $\sigma_i=\sigma$.
 We similarly define $P_{i\backslash a}(\sigma)$ as a probability of $\sigma_i=\sigma$ on a cavity graph $G\backslash a$.
 These probabilities are regarded as messages on the graph. They
 satisfy the following recursive relations: 
\begin{align}
P_{i\backslash a}(\sigma)&\simeq \frac{1}{Z_{i\rightarrow a}}\exp(-\mu \sigma-\mu^r \delta_{\sigma,0})
\prod_{b\in\partial i\backslash a}P_{b\backslash i}(\sigma), \label{eq_a31}\\
P_{a\backslash i}(\sigma)&\simeq \frac{1}{Z_{a\rightarrow i}}\sum_{\bm{\sigma}_{\partial a\backslash i}}
\theta\left(\sigma+\sum_{j\in\partial a\backslash i}\sigma_j+\alpha-2\right)\nonumber\\
\times&\prod_{j\in\partial a\backslash i}P_{j\backslash a}(\sigma_j).  \label{eq_a32}
\end{align}
 By substituting a spin value, we obtain
\begin{align}
P_{i\backslash a}(1)&\simeq \frac{1}{Z_{i\rightarrow a}}e^{-\mu}
\prod_{b\in\partial i\backslash a}P_{b\backslash i}(1), \nonumber\\
P_{i\backslash a}(0)&\simeq \frac{1}{Z_{i\rightarrow a}}e^{-\mu^r}
\prod_{b\in\partial i\backslash a}P_{b\backslash i}(0), \nonumber\\
P_{i\backslash a}(-1)&\simeq \frac{1}{Z_{i\rightarrow a}}e^{\mu}
\prod_{b\in\partial i\backslash a}P_{b\backslash i}(-1),  \label{eq_a4}
\end{align}
 and
\begin{align}
P_{a\backslash i}(1)&\simeq \frac{1}{Z_{a\rightarrow i}}, \nonumber\\
P_{a\backslash i}(0)&\simeq \frac{1}{Z_{a\rightarrow i}}
\left(1-\prod_{j\in\partial a\backslash i}P_{j\backslash a}(-1)\right)
, \nonumber\\
P_{a\backslash i}(-1)&\simeq \frac{1}{Z_{a\rightarrow i}}\left(1-\prod_{j\in\partial a\backslash i}P_{j\backslash a}(-1)\right.\nonumber\\
-&\left.\sum_{k\in \partial a\backslash i}P_{k\backslash a}(0)\prod_{j\in\partial a\backslash \{i,k\}}P_{j\backslash a}(-1)\right).
\label{eq_a5}
\end{align}

It is convenient to introduce 
cavity fields defined as shown below: 
\begin{align}
P_{i\backslash a}(\sigma)&\equiv \frac{e^{\mu\xi_{i\rightarrow a}\delta(\sigma,1)+\mu\nu_{i\rightarrow a}\delta(\sigma,0)}}
{1+e^{\mu\xi_{i\rightarrow a}}+e^{\mu\nu_{i\rightarrow a}}},\nonumber\\
P_{a\backslash i}(\sigma)&\equiv \frac{e^{\mu\hat{\xi}_{a\rightarrow i}\delta(\sigma,1)+\mu\hat{\nu}_{a\rightarrow i}\delta(\sigma,0)}}
{1+e^{\mu\hat{\xi}_{a\rightarrow i}}+e^{\mu\hat{\nu}_{a\rightarrow i}}}
,  \label{eq_a4b}
\end{align} 
where $\delta(\cdot,\cdot)$ is Kronecker's delta.
BP equations for these fields are explicitly written down as 
\begin{align}
\xi_{i\rightarrow a}&=-2+\sum_{b\in\partial i\backslash a}\hat{\xi}_{b\rightarrow i},\nonumber\\
\nu_{i\rightarrow a}&=-1-\mu^{r-1}+\sum_{b\in\partial i\backslash a}\hat{\nu}_{b\rightarrow i},\nonumber\\
\hat{\xi}_{a\rightarrow i}&=\frac{1}{\mu}\ln\left[
1-\left(1+\sum_{k\in \partial a\backslash i}e^{\mu\nu_{k\rightarrow a}}\right)\right.\nonumber\\
\times&\left.\prod_{j\in\partial a\backslash i}\frac{1}{1+e^{\mu\xi_{j\rightarrow a}}+e^{\mu\nu_{j\rightarrow a}}}\right], \nonumber\\
\hat{\nu}_{a\rightarrow i}&=\frac{1}{\mu}\ln\left[1-\prod_{j\in\partial a\backslash i}
\frac{1}{1+e^{\mu\xi_{j\rightarrow a}}+e^{\mu\nu_{j\rightarrow a}}}\right]\nonumber\\
-\frac{1}{\mu}&\ln\left[1-\left(1+\sum_{k\in \partial a\backslash i}e^{\mu\nu_{k\rightarrow a}}\right)
\prod_{j\in\partial a\backslash i}\frac{1}{1+e^{\mu\xi_{j\rightarrow a}}+e^{\mu\nu_{j\rightarrow a}}}\right]
.  \label{eq_a6}
\end{align}

 Here, we consider a graph ensemble for which the degree distribution of variable nodes is $p_k$ ($k\ge 0$).
 Letting $\tilde{P}(\xi,\nu)$ be a frequency distribution of a set of cavity fields $(\xi,\nu)$, then
 from Eq.~(\ref{eq_a6}), we find a self-consistent equation of $\tilde{P}(\xi,\nu)$ as
\begin{widetext}
\begin{align}
\tilde{P}(\xi,\nu)&=\sum_{k=0}^\infty\frac{kp_k}{c}\int\prod_{i=1}^{k-1}\prod_{j=1}^{\alpha-1}d\tilde{P}\left(\xi^{(i,j)},\nu^{(i,j)}\right)
\delta\left(\xi+2+\sum_{i=1}^{k-1}v_2\left(\{\xi^{(i,j)}\},\{\nu^{(i,j)}\};\mu\right)\right)\nonumber\\
\times& \delta\left(\nu+1+\mu^{r-1}-\sum_{i=1}^{k-1}\left[v_1\left(\{\xi^{(i,j)}\},\{\nu^{(i,j)}\};\mu\right)
-v_2\left(\{\xi^{(i,j)}\},\{\nu^{(i,j)}\};\mu\right)\right]\right),
  \label{eq_a7}
\end{align}
where
\begin{equation}
v_1\left(\{\xi^{(i,j)}\},\{\nu^{(i,j)}\};\mu\right)
=\frac{1}{\mu}\ln\left[1-
\prod_{j=1}^{\alpha-1}\frac{1}{1+e^{\mu\xi^{(i,j)}}+e^{\mu\nu^{(i,j)}}}\right],
  \label{eq_a8}
\end{equation}
and
\begin{equation}
v_2\left(\{\xi^{(i,j)}\},\{\nu^{(i,j)}\};\mu\right)
=\frac{1}{\mu}\ln\left[1-\left(1+\sum_{j}e^{\mu\nu^{(i,j)}}\right)
\prod_{j=1}^{\alpha-1}\frac{1}{1+e^{\mu\xi^{(i,j)}}+e^{\mu\nu^{(i,j)}}}\right].
  \label{eq_a9}
\end{equation}
\end{widetext}

 To obtain the single-spin probability $P_i(\sigma)$, we also introduce effective fields such as cavity fields
 and obtain the frequency distribution of those fields.
 In the case of Erd\"os--R\'enyi random graphs, the distribution is equivalent to that of cavity fields
 because an identity $p_{k-1}=kp_{k}/c$ ($k\ge 1$) holds.

 By interpreting 
 the definition of effective fields appropriately, it is apparent that the self-consistent equation is equivalent
 to Eq.~(\ref{eq_38}) obtained using the replica method.
 Further assumptions are necessary to analyze the case of the
 large $\mu$ limit. 
 They correspond to the ansatz discussed in Sec.~\ref{ssec34}.
 We correctly obtain the typical property of the LP relaxation by taking the LP-limit.

\end{document}